# How does the electric current propagate through the fully-hydrogenated borophene?

Yipeng An,*[a,b] Jutao Jiao,[a] Yusheng Hou,[b] Hui Wang,[b] Dapeng Wu,[c] Tianxing Wang,[a] Zhaoming Fu,[a] Guoliang Xu,[a] and Ruqian Wu*[b]

[a]*College of Physics and Materials Science & International United Henan Key Laboratory of Boron Chemistry and Advanced Energy Materials, Henan Normal University, Xinxiang 453007, China. Email: ypan@htu.edu.cn*

[b]*Department of Physics and Astronomy, University of California, Irvine, California 92697-4575, USA. Email: wur@uci.edu*

[c]*School of Chemistry and Chemical Engineering & International United Henan Key Laboratory of Boron Chemistry and Advanced Energy Materials, Henan Normal University, Xinxiang 453007, China*

We study the electronic transport properties of two-dimensional (2D) fully-hydrogenated borophene (namely, borophane), using the density functional theory and non-equilibrium Green's function approaches. Borophane shows a perfect electrical transport anisotropy and is promising for applications. Along the peak- or equivalently the valley-parallel direction, the 2D borophane exhibits a metallic characteristic and its current-voltage (*I-V*) curve shows a linear behavior, corresponding to the ON state in borophane-based nano-switch. In this case, electrons mainly propagate *via* the B-B bonds along the linear boron chains. In contrast, the electron transmission is almost forbidden along the perpendicular buckled direction

(*i.e.*, the OFF state), due to its semi-conductor property. Our work demonstrates that 2D borophane could combine the metal and semiconductor features and can be a promising candidate of nano-switching materials with stable structure and high ON/OFF ratio.

# 1 Introduction

Borophene is a new atomically thin two-dimensional nano-sheet and has been successfully synthesized on metallic (*e.g.*, silver) substrates under the ultrahigh-vacuum condition.[1,2] Due to its unique geometric and electronic structures, borophene has attracted broad attention in recent years and many possible exploitations in innovative devices have been proposed.[3-5] For instance, Xiang *et al.*[3] studied the metallic borophene polytypes as lightweight anode materials for non-lithium-ion batteries. Their results provide insights into the configuration-dependent electrode performance of borophene and the corresponding metal-ion storage mechanism. Kistanov *et al.*[4] investigated the routes to localize the carriers and open the band gap of borophene *via* several methods, such as chemical functionalization, ribbon construction, and defect engineering. Jian *et al.*[5] explored the potential of using borophene and defective borophene as anchoring materials for Li-S batteries on the basis of first-principles computations. They found that the defective borophene could be a promising anchoring material and lead to a suppressed shuttle effect and enhanced capacity retention for Li-S batteries.

However, 2D borophene is unstable without suitable substrate based on the theoretical calculations.[6,7] One viable method to stabilize borophene is surface hydrogenation. Xu *et al.*[6] studied the stability of the fully-hydrogenated borophene

(borophane) using the first-principles calculation and found that borophane might be stable in vacuum without a substrate. Furthermore, borophane has an ultrahigh Fermi velocity, 4 times higher than that of graphene. Another important feature of borophene is its high mechanical anisotropy along two different directions. Jena *et al.*[7] investigated the possibilities of using borophane as a potential Li and Na-ion battery anode material through the first-principles calculations. Nevertheless, the electronic and transport properties of borophane have not been carefully examined, which somewhat hindered its applications in nano-electronic devices.

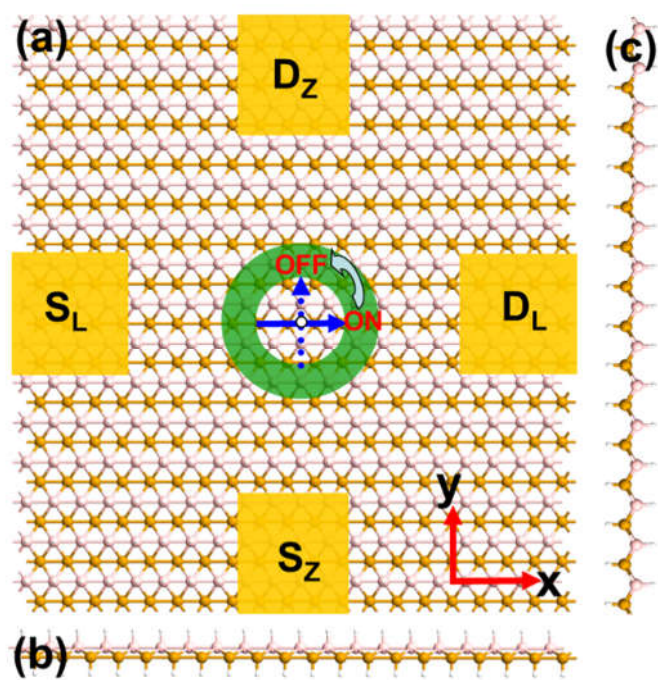

**Fig. 1** The nano-switch structure based on the fully-hydrogenated borophene. (a) top, (b) and (c) side views. $S_{L/Z}$ and $D_{L/Z}$ refer to the source and drain electrodes of two-probe system along the *x/y* direction. The pink/golden color refers to the boron atoms at peak/valley position.

Borophane has a much better stability than borophene. The main issues regarding its electronic properties are: (1) how strong is its electrical anisotropy? (2) how does the electrical current propagate through the 2D borophane? (3) what unique feature(s) does it have for the use in nano-electronics? To answer these questions, we systematically study the electronic transport properties of borophane (Fig. 1) through density functional theory (DFT) calculations. We find that the 2D borophane has anisotropic current-voltage characteristics along two orthogonal directions, namely, along the *x* axis (labeled as the L type) and along the *y* axis (labeled as the Z type), respectively. The *I-V* curve of the L type displays a perfect linear behavior, while it remains zero for the Z type. This suggests that the transports properties of borophane strongly depends on the pattern of hydrogenation and thus can be manipulated for the uses as nano-sensing and nano-switching.

**2 Sample structure and method**

Figure 1 shows the sample structure of the fully-hydrogenated borophene's two-probe systems, including the top (Fig. 1a) and side views (Figs. 1b and c). We explore its electrical properties along the two orthogonal directions. Namely, one is along the *x* axis, and the other is along the *y* axis. In the two-probe structure for each type (L or Z), a 2D supercell is constructed with a periodicity perpendicular to the direction of current between the source ($S_{L/Z}$) and drain ($D_{L/Z}$) electrodes. The third direction (*i.e.*, *z* axis) is out of the plane, along which the slabs are separated by a 30 Å vacuum. Both $S_{L/Z}$ and $D_{L/Z}$ are semi-infinite in length along the transport direction and are described by a large supercell.

The electronic structures and transport properties are determined by using the first-principles density functional theory and the non-equilibrium Green's function method (*i.e.* NEGF–DFT).[8-11] The GGA-PBE functional is adopted to describe the exchange correlation interaction among electrons.[12,13] For all boron and hydrogen atoms, their core electrons are described by the optimized norm-conserving Vanderbilt (ONCV) pseudo-potentials, and wave functions of valence states of outer electrons are expanded as linear combinations of atom orbitals (LCAO), based on the level of SG15[14] pseudo-potentials and basis sets. Note that the SG15 datasets of ONCV pseudo-potentials are fully relativistic and provide a good agreement with the all-electron results.[14] For the structural optimization, the residual force on each atom is no more than $10^{-6}$ eV/Å. We use a 1×9×150 Monkhorst-Pack *k*-points grid to sample the Brillouin zone of the left and right electrodes. Note that in the transport direction, the self-energy calculation effectively corresponds to an infinite number of *k*-points. In this work, we use a 150 *k*-mesh to achieve a balance between the accuracy and cost, and the electronic structures of the electrodes and central scattering region match well at zero-bias with this selection. Meanwhile, we use 9 *k*-points in the periodic direction.

**3 Results and discussions**

Fig. 2a depicts the transmission spectra of the L and Z type two-probe structures of the fully-hydrogenated borophenes. One can see that both of them exhibit the quantization step characteristics, similar to that of graphene or other graphene-like structures.[15-17] Remarkably, the transmission spectrum of the L type exhibits a platform near the Fermi level ($E_F$), while it is almost zero for the Z type near the $E_F$.

Such remarkable difference should bring about distinct current-voltage behaviors when the bias is applied to the source and drain electrodes.

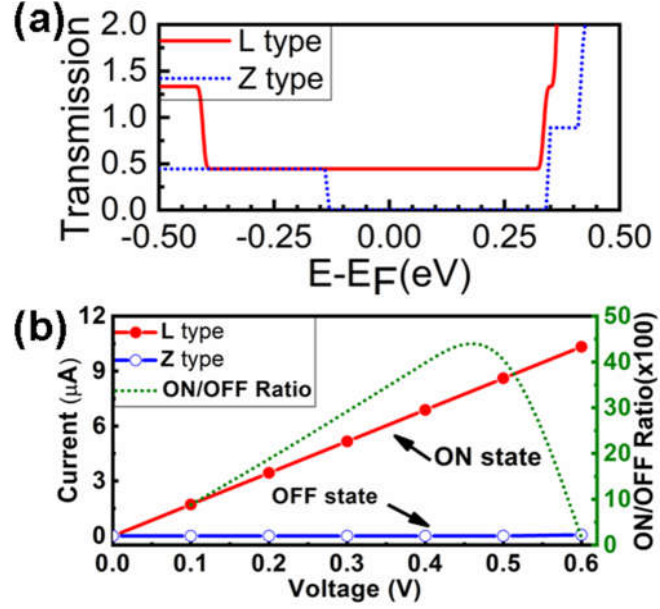

**Fig. 2** (a) Transmission spectra of L and Z type borophane two-probe structures under zero bias, (b) their I-V and ON/OFF ratio curves.

When a bias $V_b$ is applied, the energies of the source and drain electrodes are shifted accordingly. A positive bias gives rise to an electric current from the source to the drain, and vice versa. In the present work, the current $I$ through the L and Z type borophane two-probe systems is obtained by using the Landauer–Büttiker formula[18]

$$I(V_b) = \frac{2e}{h}\int_{-\infty}^{\infty} T(E,V_b)[f_S(E-\mu_S) - f_D(E-\mu_D)]dE, \quad (1)$$

where $T(E,V_b)$ is the bias-dependent transmission coefficient, calculated from the Green's functions, $f_{S/D}$ is the Fermi-Dirac distribution function of the left (L)/right (R) electrode. $\mu_S$ (= $E_F - eV_b/2$) and $\mu_D$ (= $E_F + eV_b/2$) are the electrochemical potentials of the source and drain electrodes, respectively. More details of the method can be found in previous publications.[10,11]

Fig. 2b gives the *I-V* curves of L and Z type borophane two-probe structures. We can see that they indeed exhibit obvious difference under the limited biases from 0 to 0.6 V. This indicates that borophane has an outstanding electrical anisotropy property, in addition to its mechanical anisotropy property.[6] More explicitly, the *I-V* curve of L type displays a perfect linear behavior like metals. However, the electric current through the Z type is almost zero like semi-conductors or insulators. Based on such *I-V* behaviors of the L and Z type, we propose that the borophane could be used as a nano-switch. That is, the L type is the ON state of the nano-switch, while the Z type corresponds to its OFF state. Then we further calculate its ON/OFF ratio (equals to $I_L/I_Z$), as shown in Fig. 2b. One can see that such borophane nano-switch has large ON/OFF ratios under the limited biases, and has the peak value at the bias of 0.5 V, which is about 5000 and larger than that of a single-molecule photoswitch (~100).[19]

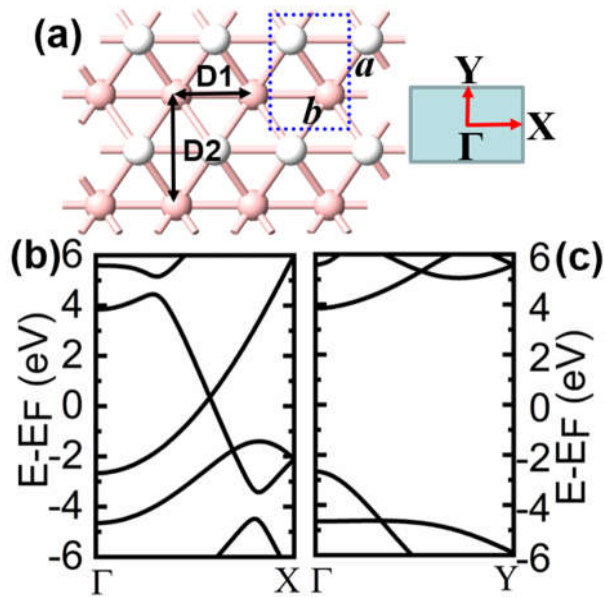

**Fig. 3** The super cell (a) and band structures of borophane along *x* (b) and *y* (c) directions. The Fermi level is set to zero. D1/D2 refers to the distance between the nearest two boron atoms along the transport direction *x*/*y*. The blue dashed box in (a)

refers to the unit cell of the borophane, and the first Brillouin zone is shown in the right side.

To understand the large anisotropy of the electronic transport of borophane, we firstly analyze the geometric structure of borophane. The fully-hydrogenated borophene is still out-of-plane buckled and has a similar wrinkle with that of borophene.[1] The blue dashed box in Fig. 3a shows the unit cell of the borophane. We use a 99×99 *k*-points grid for its geometry optimization and band structures calculation. The optimized lattice constants *a* (= D1) and *b* (= D2) were determined to be 2.82 and 1.93 Å, agreeing well with previous investigations.[6,20] Along the two different directions, *x* and *y*, the distance (D1/D2) between the nearest boron atoms has obvious difference. As shown in Fig. 3a, for the L type, the D1 (1.93 Å) is far smaller than that (D2) of the Z type, which is about 2.82 Å. Importantly, the nearest boron atoms along the transport direction bind together for the L type, but not for the Z type.

Moreover, the electronic transport properties should depend on their band structures. Namely, electron transmission contains contributions from intra- and inter-band transitions around the Fermi level. To unveil the physical origin of the anisotropy of the electronic transport of borophane, we calculate its band structures along *x* and *y* directions, which is located on the *k*-lines from $\Gamma$ to $X$ ($\Gamma$–$X$) and from $\Gamma$ to $Y$ ($\Gamma$–$Y$), respectively. It shows completely different band structures along the two *k*-lines $\Gamma$–$X$ and $\Gamma$–$Y$ (see Figs. 3b and c). One may see that the band structures along $\Gamma$–$X$ line (for the L type transport direction) shows a perfect metallic characteristic like that of 2D borophene.[1] However, it presents a semiconductor property with a large direct band gap (~6.5 eV) along $\Gamma$–$Y$ line (for the Z type transport direction). All of the results are consistent with previous studies.[6,21] Thus, the anisotropy of the electronic transport of borophane is mostly attributed to the intrinsic difference between its band structures along the $x$ and $y$ directions.

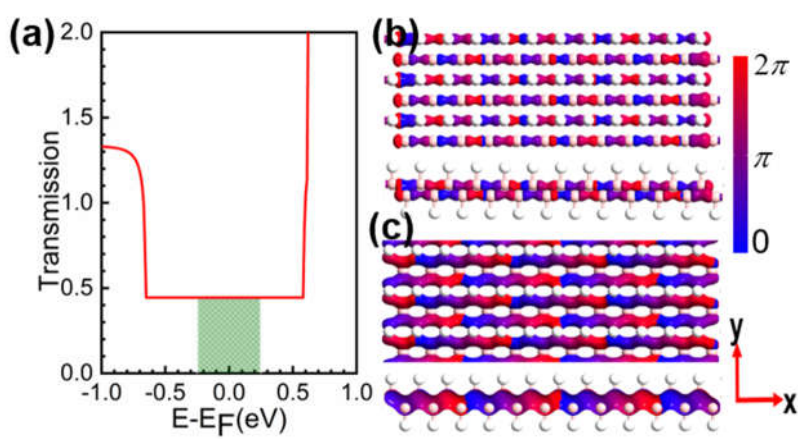

**Fig. 4** (a) Transmission spectrum under the bias of 0.5 V, (b) and (c) refer to the two different degenerate transmission eigenstates (TE-I and TE-II) at the Fermi level, including the top and side views. The shadow in (a) refers to the bias window, which contributes to the electric current.

To further disclose the transport channel of borophane at the atomic level, we investigate the transmission spectrum of the L type under the bias of 0.5 V (Fig. 4a) in detail. One can observe that the wide platform is still kept, and the linear electric current is produced near the $E_F$. There are two different degenerate transmission eigenstates (*i.e.*, TE-I and TE-II) at the $E_F$, as shown in Figs. 4b and c. Specifically, the TE-I is composed of $p_x$ orbitals of boron atoms, and almost uniformly spread over each boron chain along the $x$ direction. Such continuously uniform distribution is beneficial for electron transmission, and mostly makes the borophane nano-switch open with high mobility. For the TE-II, however, it is disrupted across boron chains and hence the electron transmission along the *y* axis is effectively blocked.

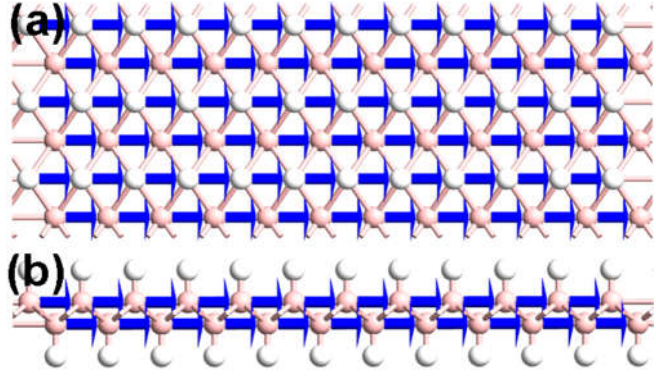

**Fig. 5** (a) Top and (b) side views of local transmission pathways of L type two-probe structures at the $E_F$ under the bias of 0.5 V.

To give an intuitive understanding on the electron transmission pictures of borophane, we plot the transmission pathways (namely, local currents)[21] of the central scattering region of the L type two-probe structure at the $E_F$ under the bias of 0.5 V (see its top and side views in Figs. 5a and b). The electron transmission pathway is an analysis option that splits the transmission coefficient into local bond contributions $T_{ij}$. It has the property that if the system is divided into two parts (A, B), then the pathways across the boundary between A and B sum up to the total transmission coefficient

$$T(E)=\sum_{i\in A, j\in B} T_{ij}(E) \quad . \tag{2}$$

Generally speaking, there are two types of local current pathways under a bias. One is *via* chemical bonds (*i.e.*, bond current), and the other is through electron hopping (*i.e.*, hopping current) between atoms which have no strong bonding.[22,23] Our calculated results (see Fig. 5) show that there only exists the bond current for the L type borophane. This bond current locates along the nearest B-B bonds in boron chain, including both the peak and valley positions of the wrinkled fully-hydrogenated borophene. Such linear pathways should have the least barrier and electron scattering for electron transmission. The linear electron transmission pathways are mostly attributed to the TE-I transmission eigenstates, while little contributed by the TE-II transmission eigenstates, as shown in Figs. 4b and c.

## 4 Conclusions

In conclusion, by using the first-principles density functional theory and the non-equilibrium Green's function method, we investigate the electronic transport properties of the fully-hydrogenated borophene, which still keeps a buckled structure but is more stable. Our results demonstrate that borophane displays a huge electrical anisotropy along with the mechanical anisotropy. Such peculiar electrical anisotropy makes borophane a promising candidate of nano-switches with high ON/OFF ratio. In the peak- and valley-parallel direction (*i.e.*, the L type two-probe structure along the *x* axis), it shows metallic characteristic and its *I-V* curve is linear. It may keep the circuit conductive, corresponding to the ON state of the borophane-based nano-switch. The $p_x$ orbitals of boron atoms along the *x* axis mediate the electron transmission through the B-B bonds. Along the buckled direction, in contrast, borophane shows a semiconductor property, corresponding to the OFF state of the borophane-based nano-switch. The giant electrical anisotropy of the 2D fully-hydrogenated borophene is sensitive to the order of H atoms and thus can be useful for nano-sensing.

**Acknowledgements**

Work at HNU was supported by the National Natural Science Foundation of China (Grant Nos. 11774079, U1704136 and 51772078), the Natural Science Foundation of Henan Province (Grant No. 162300410171, 152300410085), the CSC (Grant No. 201708410368), the Science Foundation for the Excellent Youth Scholars of Henan Normal University (Grant No. 2016YQ05), and the High-Performance Computing Centre of Henan Normal University. Work at UCI was supported by DOE-BES (Grant No. DE-FG02-05ER46237).